\documentclass[3p,times,procedia]{elsarticle}

\UseRawInputEncoding
\usepackage{float}
\usepackage{amsmath,amssymb}
\usepackage{bm}
\usepackage{slashed}
\usepackage{braket}
\usepackage{epsfig}
\usepackage{graphicx}
\usepackage{hyperref}
\usepackage{xcolor}
\usepackage{subcaption}
\hypersetup{
    colorlinks=true,
    linkcolor=blue,
    filecolor=magenta,      
    urlcolor=blue,
    citecolor=blue
}
\newcommand{\comment}[1]{}

\begin{document}
\title{Quark production in the bottom-up thermalization}
\author{Sergio Barrera Cabodevila}
\author{Xiaojian Du}
\author{Carlos A. Salgado}
\author{Bin Wu}
\address{Instituto Galego de F\'isica de Altas Enerx\'ias IGFAE, Universidade de Santiago de Compostela, E-15782 Galicia-Spain}
\begin{abstract}
We investigate the impact of quark production on bottom-up thermalization in heavy-ion collisions. First, we extend the parametric estimates of bottom-up thermalization in pure gluon systems by incorporating quark production in the weak-coupling (high-energy) limit. Our analysis reveals that quark production does not alter the qualitative features of the three-stage thermalization process in this limit. Furthermore, we obtain the scaling behavior of the quark number density over time at each stage. Then, by solving the Boltzmann equation in diffusion approximation (BEDA) for longitudinally boost-invariant systems, we demonstrate how our detailed numerical simulations approach the predicted three-stage thermalization picture as the strong coupling $\alpha_s$ decreases. Finally, we carry out a detailed comparison of our BEDA results with those obtained by solving the QCD effective kinetic theory for intermediate values of $\alpha_s$, observing remarkably good quantitative agreement between the two approaches.
\end{abstract}

\maketitle

\section{Introduction} 

One of the most fundamental unresolved theoretical questions in heavy-ion physics is to understand, from QCD first principles, how small droplets of quark-gluon plasma (QGP) emerge in high-energy nucleus-nucleus collisions~\cite{Schlichting:2019abc,Berges:2020fwq}. While no reliable framework currently exists to describe this process at intermediate or large coupling, a consistent picture is beginning to emerge in the weak-coupling (high-energy) regime within perturbative QCD. According to the concept of parton saturation~\cite{Mueller:2001fv}, saturated partons, characterized by the saturation momentum $Q_s$, are freed from the colliding nuclei, giving rise to a dense partonic system predominantly composed of gluons in the very early stage of the collision~\cite{Jalilian-Marian:1996mkd, Kovchegov:1998bi, Mueller:1999fp}. The saturation momentum $Q_s$ at LHC energies is estimated to be about 1-2~GeV, entering the perturbative regime.

In the weak-coupling limit, the subsequent thermalization process of pure gluon systems has been outlined in Ref.~\cite{Baier:2000sb} through parametric estimates, known as the "bottom-up" thermalization. Quantitative studies have been investigated using weak-coupling techniques, including classical statistical field theory~\cite{Berges:2013eia, Epelbaum:2013ekf} and effective kinetic theory (EKT)~\cite{Arnold:2002zm}, as presented in~\cite{Kurkela:2015qoa, Kurkela:2018xxd, Kurkela:2018oqw, Du:2020dvp}. Classical field approaches are known to break down at later times~\cite{Mueller:2002gd, Berges:2013lsa, Epelbaum:2014yja, Epelbaum:2014mfa}. It is conventional to switch from classical field simulations to the Boltzmann equation description~\cite{Baier:2000sb, Mueller:2002gd, Kurkela:2018vqr}, although how this transition can be understood in perturbation theory remains unclear~\cite{Wu:2017rry, Kovchegov:2017way}. To date, EKT has been the only weak-coupling framework used to study quantitatively the entire process of subsequent thermalization/hydrodynamization.

Recently, the Boltzmann equation in diffusion approximation (BEDA) has been solved to explore various aspects of thermalization in spatially homogeneous systems~\cite{BarreraCabodevila:2022jhi, Cabodevila:2023htm}. This set of equations includes both $2\leftrightarrow2$ interactions~\cite{Mueller:1999pi, Baier:2000sb, Hong:2010at, Blaizot:2013lga, Blaizot:2014jna} and 
$1\leftrightarrow2$ processes due to the Landau-Pomeranchuk-Migdal (LPM) effects in the deep LPM regime~\cite{Baier:1998kq, Arnold:2008zu}. Using BEDA, parametric estimates for thermalization have been derived for both initially under- and over-populated systems and verified through numerical solutions. Moreover, the results remain qualitatively consistent with those from EKT~\cite{AbraaoYork:2014hbk, Kurkela:2014tea, Kurkela:2018oqw, Kurkela:2018xxd, Du:2020dvp, Du:2020zqg}, while benefiting from reduced computational complexity due to the diffusion approximation.

As elaborated in Refs.~\cite{Kurkela:2014tea, BarreraCabodevila:2022jhi, Cabodevila:2023htm}, spatially homogeneous, initially over-populated systems exhibit similarities to the early stage of bottom-up thermalization. This stage is characterized by self-similar solutions, referred to as a nonthermal fixed point in Ref.~\cite{Berges:2013eia}. The characteristic scaling behavior can be understood via momentum broadening due to multiple elastic scatterings~\cite{Baier:2000sb}, which can be alternatively studied via the adiabatic hydrodynamization approach~\cite{Brewer:2019oha, Rajagopal:2024lou}. On the other hand, initially under-populated systems without expansion also undergo bottom-up thermalization in the final stage of thermalization, driven by democratic branching due to the LPM effects~\cite{Baier:2000sb, Baier:2001yt, Blaizot:2013hx, Iancu:2015uja}. Allowing quark production does not significantly alter the qualitative behavior of initially very dilute or dense systems~\cite{Cabodevila:2023htm}. However, in non-extreme cases, the system thermalizes in a top-down manner: gluons thermalize first at a higher temperature and gradually cool down to the final equilibrium temperature as quarks are produced~\cite{Kurkela:2018oqw, Cabodevila:2023htm}.

In longitudinally boost-invariant systems, quark production has been recently investigated  using EKT~\cite{Kurkela:2018oqw, Kurkela:2018xxd, Du:2020dvp, Du:2020zqg}. Such theoretical studies are of significance to phenomenological investigations of pre-equilibrium photon~\cite{Garcia-Montero:2023lrd} and di-lepton production~\cite{Coquet:2021lca,Coquet:2021gms,Coquet:2023wjk,Garcia-Montero:2024lbl}, as well as heavy quark thermalization~\cite{Du:2023izb} at the early stages of heavy-ion collisions. In this work, we mainly focus on the role played by quark production during the bottom-up thermalization in the weak-coupling limit, providing both parametric estimates and numerical validation. Additionally, we conduct a detailed comparison between BEDA and EKT solutions at intermediate coupling, highlighting their quantitative similarities.

\section{The BEDA for longitudinally boost-invariant systems}

For a system that is longitudinally boost-invariant and uniform in the transverse plane~\cite{Bjorken:1982qr}, the QCD Boltzmann equation at midrapidity ($z=0$) takes the form~\cite{Mueller:1999pi}:  
\begin{align}  
\left(\partial_{\tau} - \frac{p_z}{\tau} \partial_{p_z} \right) f^a = C^a_{2\leftrightarrow2} + C^a_{1\leftrightarrow2},  
\end{align}  
where $f^a$ denotes the distribution function of parton $a$, which varies with time (equal to proper time $\tau\equiv\sqrt{t^2-z^2}$), momentum magnitude $p$, and longitudinal velocity $v_z = p_z/p$. Our analysis focuses on the impact of quark production on the entire bottom-up thermalization process, assuming the absence of initial quarks and antiquarks. Consequently, the distribution functions for all $N_f$ massless quark (and antiquark) flavors remain identical, as the Boltzmann equation preserves both $q \leftrightarrow \bar{q}$ symmetry and flavor symmetry. And the effective net quark chemical potentials all vanish. For brevity, we denote the distribution functions for different species as
\begin{align}
f = f^g, \qquad F = f^q = f^{\bar{q}}.
\end{align}

As recently detailed in~\cite{Cabodevila:2023htm}, the $2\leftrightarrow2$ kernel is simplified using the diffusion approximation~\cite{Mueller:1999pi, Baier:2000sb, Hong:2010at, Blaizot:2013lga, Blaizot:2014jna}: 
\begin{align}\label{eq:C2to2}
C^a_{2\leftrightarrow2}=\frac{1}{4}\hat{q}_a(t)\nabla_{{\bm p}} \cdot\left[  \nabla_{{\bm p}}f^a  + \frac{{\bm v}}{T^*(t)}f^a(1+\epsilon_a f^a)\right]+\mathcal{S}_a,
\end{align}
where $\epsilon_a=1$ for bosons and  $\epsilon_a=-1$ for fermions and the source terms~\cite{Blaizot:2014jna} take the form
\begin{align}\label{eq:Sqg}
   \mathcal{S}_{q} =\frac{2\pi\alpha_s^2  C_F^2 \mathcal{L}}{ p}\mathcal{I}_c\bigg[ f (1 - F ) - F ( 1 + f ) \bigg],\qquad
   \mathcal{S}_g=-\frac{N_f}{C_F}\mathcal{S}_{q}.
\end{align}
The space-time-dependent quantities in the above equations, which govern the time evolution of the system, are defined as:  

1) The jet quenching parameter is defined as $\hat{q}_a \equiv C_a \hat{\bar{q}}$~\cite{Baier:1996sk}, where 
\begin{align}
    \hat{\bar{q}} \equiv 8\pi \alpha_s^2\mathcal{L}\int\frac{d^3\bm p}{(2\pi)^3}
    \left[N_c f (1+f) + N_f F (1 - F) \right].
\end{align}  
Here, $C_a$ takes the values $C_A = N_c$ for gluons and $C_F=(N_c^2-1)/(2N_c)$ for quarks/antiquarks. The logarithmic factor is defined as $\mathcal{L} \equiv \ln({\langle p_t^2\rangle}/{m_D^2})$. Different choices of the typical transverse momentum broadening $\langle p_t^2\rangle$ in $\mathcal{L}$ only introduce corrections beyond the leading logarithmic approximation. In this work, it is taken as $\langle p_t^2\rangle = \hat{\bar{q}} t_{\text{br}}(\bar{p})$, where
the branching time is given by $t_{\text{br}}(p)={\sqrt{p/\hat{\bar{q}}}}/{\alpha_s}$~\cite{Iancu:2015uja}. This leads to
\begin{align}
\label{eq:Lchoice}
    \mathcal{L} = \ln \bigg(\frac{\hat{\bar{q}} t_{\text{br}}(\bar{p})}{m_D^2}\bigg) = \ln \bigg(\frac{\sqrt{\hat{\bar{q}}\bar{p}}}{\alpha_s m_D^2}\bigg),
\end{align}  
where the typical momentum $\bar{p}$ is defined as the square root of the average momentum square $p^2$ per parton, i.e., $ \bar{p} =\sqrt{\langle p^2 \rangle}$, and the logarithmic dependence of $\hat{\bar{q}}$ in the argument of the log on the right-hand side of Eq.~(\ref{eq:Lchoice}) is ignored. Under this choice, in a thermalized QGP, we have $\mathcal{L}\sim \ln(1/\alpha_s)$.

2) The effective temperature $T_*$ is defined as  
\begin{align}\label{eq:Ts}
    T_* \equiv \frac{\hat{q}_A}{\alpha_s N_c \mathcal{L} m_D^2},
\end{align}  
where the screening mass squared is given by  
\begin{align}
    m_D^2 \equiv 16\pi\alpha_s\int\frac{d^3 \bm{p}}{(2\pi)^3}\frac{1}{p} (N_c f + N_f F).
\end{align}  

3) The conversion coefficient, which determines the rate of quark-gluon conversion in $2\leftrightarrow2$ scatterings~\cite{Blaizot:2014jna}, is expressed as
\begin{align}\label{eq:IcIcb}
    \mathcal{I}_c = \int \frac{d^3\bm{p}}{(2\pi)^3} \frac{1}{p} (f + F).
\end{align}

The 
$1\leftrightarrow2$ kernel includes all the collinear splittings:
\begin{align}
    C^a_{1\leftrightarrow2} = &\int_0^1 dx\sum\limits_{b,c}\bigg[\frac{1}{x^3}\frac{\nu_c}{\nu_a}C^c_{ab}({\bm p}/{x};{\bm p},{\bm p}(1-x)/x)-\frac{1}{2}C^a_{bc}(\bm p;x{\bm p},(1-x){\bm p})\bigg],
    \label{eq:inel_kern}
\end{align}
where $x$ presents the momentum fraction carried by particle $b$ for the process $a\to\,bc$, and $\nu_a$ is the number of spin times color degrees of freedom for parton $a$: $\nu_q=2 N_c$ for quarks (antiquarks) and $\nu_g=2 (N_c^2-1)$ for gluons. Here, we define
\begin{align}
    C^{a}_{bc}(\bm p;x{\bm p},(1-x){\bm p})\equiv\frac{dI_{a\to bc}(p)}{dxdt}\mathcal{F}^a_{bc}(\bm p;x{\bm p},(1-x){\bm p}),
    \label{eq:inel_kern2}
\end{align}
and
\begin{align}
    \mathcal{F}^a_{bc}(\bm p;{\bm l},{\bm k})\equiv f^a_{\bm p}(1+\epsilon_b f^b_{\bm l})(1+\epsilon_c f^c_{\bm k})-f^b_{\bm l}f^c_{\bm k}(1+\epsilon_a f^a_{\bm p})
    \label{eq:inel_kern3}
\end{align}
with ${\bm l}\equiv x {\bf p}$ and ${\bm k}\equiv (1-x){\bf p}$. The splitting rates ${dI_{a\to bc}(p)}/{dxdt}$ account for the LPM effects
in the deep LPM regime~\cite{Baier:1998kq, Arnold:2008zu}. Their exact expressions in terms of the jet quenching parameter are compiled in Ref.~\cite{Cabodevila:2023htm}. Here, we highlight the parametric difference between the splittings $a\to g a$ and $g\to q\bar{q}$:
\begin{align}
\label{eq:g2qqb}
    x \frac{dI_{a\to ga}(p)}{dxdt} \approx \frac{\alpha_s C_a}{\pi} \sqrt{\frac{\hat{q}_A}{ x p}}\qquad\text{v.s.}\qquad x \frac{dI_{g\to q\bar{q}}(p)}{dxdt} \approx \frac{\alpha_s}{4\pi} \sqrt{\frac{x\hat{q}_F}{p}}
\end{align}
at $x\ll 1$. This leads to a qualitative difference between gluon and quark production via the $1\to2$ processes, as discussed below.

\section{Quark production in the bottom-up thermalization}
\label{sec:quarkprod}

In this section, we study the impact of quark production on the bottom-up thermalization in the weak-coupling limit, based on parametric estimates and numerical simulations in BEDA.

\subsection{\bf Parametric estimates}
Following Ref.~\cite{Cabodevila:2023htm}, we consider quark production through both the $2\to2$ conversion
($g\to q/\bar{q}$) and the $1\to 2$ splitting ($g\to q\bar{q}$). Starting with a generic initial gluon distribution $f\sim 1/\alpha_s$ for $p_z\sim p_{\perp}\sim Q_s$ at $\tau\sim 1/Q_s$, the system evolves  through three different stages before transitioning to Bjorken hydrodynamics~\cite{Bjorken:1982qr}:

{1.\it  Very early stage: diluting an over-occupied system during $1 \lesssim  Q_s\tau  \lesssim \alpha_s^{-{3}/{2}}$.}  During this stage, the system is primarily governed by hard gluons with momentum $\sim Q_s$, similar to pure gluon systems~\cite{Baier:2000sb}. Its main qualitative features can be deduced from the competition between expansion and multiple $2\leftrightarrow2$ scatterings. The number density of hard gluons decreases with time, following $n_{h,g}\sim {Q_s^2}/(\alpha_s\tau)$. Their longitudinal momentum undergoes broadening by an amount $\sim \sqrt{\hat{q}\tau}$ within a time interval $\sim \tau$. If this broadening is initially negligible compared to their typical longitudinal momentum $p_z$, then 
$p_z$ decreases as $1/\tau$ until a balance is reached, yielding~\cite{Baier:2000sb}:
\begin{align}
    p_z^2 \sim \hat{q} \tau \sim \alpha_s^2 \frac{n^2_{h, g}}{p_z p_\perp^2}  \tau
    \Leftrightarrow p_z\sim Q_s ( Q_s\tau )^{-{1}/{3}},
\qquad
f_{h,g}\sim \frac{n_{h,g}}{p_z p_{\perp}^2}\sim \frac{1}{\alpha_s}\frac{1}{( Q_s\tau )^{{2}/{3}}},
\qquad
\hat{q}\sim \frac{Q_s^3}{( Q_s\tau )^{{5}/{3}}}.
\end{align}
Here and below, we treat the logarithmic factor $\mathcal{L}$ as an $O(1)$ coefficient and we denote $\hat{q} \sim \hat{q}_A \sim \hat{q}_F$, since the only difference between $\hat{q}_A$ and $\hat{q}_F$ is some parametrically negligible color factor. The Debye mass squared, effective temperature, and conversion coefficient are estimated as:
\begin{eqnarray}
  m_D^2
  \sim \alpha_s \frac{n_{h,g}}{Q_s}\sim \frac{Q_s}{\tau},\qquad T_* \sim \frac{\hat{q}}{\alpha_s m_D^2}\sim \frac{Q_s}{\alpha_s ( Q_s\tau )^{2/3}},
  \qquad \mathcal{I}_c\sim m_D^2/\alpha_s.
\end{eqnarray}
The number density of soft gluons with momentum $p_s \sim p_z$, produced via the $1\to 2$ splitting, is esimated as follows:
\begin{align}
     n_{s, g} \sim \alpha_s \sqrt{\frac{\hat{q}}{p_s}} n_{h,g} f_{h,g} \tau \sim \frac{Q_s^3}{\alpha_s ( Q_s\tau )^{4/3}}\sim T_* p_s^2,\qquad f_{s, g}{\sim\frac{n_{s,g}}{p_s^3}}\sim \frac{1}{\alpha_s ( Q_s\tau )^{1/3}}
\end{align}
which fills a thermal distribution with temperature given by $T_*$ up to $p\sim p_s$~\cite{BarreraCabodevila:2022jhi}. Since $T_*$ is initially at its maximum,  the soft sector resembles that of an overheated system, similar to what is observed in spatially homogeneous, initially very dense systems~\cite{BarreraCabodevila:2022jhi, Cabodevila:2023htm}. Note that our estimates for soft gluons during this stage and the next, derived solely from the LPM effects, give the same parametric results as those obtained using the BH rate in Ref.~\cite{Baier:2000sb}.

The above estimates remain valid regardless of whether quark production is taken into account. One can estimate the quark number density produced via conversion in Eq.~(\ref{eq:Sqg}) and splitting in Eq.~(\ref{eq:g2qqb}):
\begin{align}
    n_q\sim\left\{
    \begin{array}{ll}
          m_D^4 \tau \sim (Q_s\tau)^{-1}Q_s^3& \text{for $g \to q/\bar{q}$} \\
          \alpha_s\sqrt{\hat{q}/Q_s} n_{h, g}\tau \sim (Q_s\tau)^{-5/6} Q_s^3
          & \text{for $g \to q\bar{q}$}
    \end{array}
    \right.
    .
\end{align}
Here, we have used the fact that $\alpha_s\mathcal{I}_c\sim m_D^2$, as gluons dominate. Comparing the two parametric forms, one can conclude that quark production is primarily driven by the $1\leftrightarrow2$ process at this stage, leading to  
\begin{align}
    n_q\sim ( Q_s\tau )^{-5/6} Q_s^3,
\end{align}
which is parametrically smaller than both $n_{h,g}$ and $n_{s,g}$.
These quarks mostly carry hard momentum $\sim Q_s$ while the number density of soft quarks with soft momentum $p_s\sim p_z$ is parametrically smaller, scaling as $\sim p_z^3\sim Q_s^2/\tau$. Consequently, quarks play a negligible role in determining the overall properties of the system at this stage.

Since quarks are absent at initial time ($ Q_s\tau  \sim 1$), one can directly perform the integration in the collision kernels in Eq.~(\ref{eq:inel_kern}) to examine the transition from the initial distribution to the above scaling and obtain
\begin{align}
    n_q \sim \int_{Q_s^{-1}}^\tau d\tau' (\tau' Q_s)^{-11/6}Q_s^4 \sim \frac{ Q_s\tau  - 1}{( Q_s\tau )^{11/6}} Q_s^3,
\end{align}
This result suggests a rapid initial increase in quark number density, reaching a peak shortly after the initial time. Subsequently, it decreases throughout the remainder of this stage.

Accordingly, throughout this stage, corrections to $\hat{q}$ and $m_D^2$ from soft gluons and quarks do not modify their parametric forms derived from hard gluons. And the pressure anisotropy, dominated by hard gluons as well, scales as $P_L/\epsilon\sim p_z^2/Q_s^2 \sim ( Q_s\tau )^{-2/3}$. Note that during this stage the contribution of soft gluons to the screening mass, given by $m_D^2 \sim \alpha_s n_{s,g}/p_z \sim Q_s/\tau$, remains parametrically comparable to that of hard gluons. This holds until the end of this stage at $ Q_s\tau \sim \alpha_s^{-3/2}$ when $f_{h,g}\sim \alpha_s^{-1}( Q_s\tau )^{-2/3}
\sim 1$.

{2. \it Setting up the stage for thermalization: cooling and overcooling of soft sector during $\alpha_s^{-3/2} \lesssim  Q_s\tau  \lesssim \alpha_s^{-5/2}$}. Primarily driven by expansion, the number density of hard gluons continues to decrease as $n_{h,g} \sim Q_s^2/(\alpha_s \tau)$, since they have not yet suffered significant energy loss. At $ Q_s\tau  \gtrsim \alpha_s^{-3/2}$, a key change for hard gluons is that $f_{h,g}$ falls below unity. This modifies the scaling of the jet transport parameter to:
\begin{align}
\hat{q} \sim \alpha_s^2 n_{h,g} \sim \alpha_s Q_s^2/\tau
\Rightarrow
p_z \sim \sqrt{\hat{q} \tau} \sim \alpha_s^{1/2} Q_s.
\end{align}
The soft gluon number density with soft momentum $p_s\sim p_z$ is then given by  
\begin{align}
    n_{s,g} \sim \alpha_s \sqrt{\frac{\hat{q}}{p_s}} n_{h,g} \tau \sim \alpha_s^{1/4} ( Q_s\tau )^{-1/2} Q_s^3,
    \qquad
    f_{s,g}\sim \frac{n_{s, g}}{p_s^3}\sim  \alpha_s^{-5/4} ( Q_s\tau )^{-1/2}.
\end{align}  
Consequently, the contribution of soft gluons to $\hat{q}$ remains parametrically smaller than that of hard gluons until $ Q_s\tau  \sim \alpha_s^{-5/2}$ when $n_{s,g}\sim n_{h,g}$. However, soft gluons dominate the contribution to the Debye mass since $n_{h,g}/Q_s\ll n_{s,g}/p_s$, leading to:
\begin{align}
    m_D^2 \sim \alpha_s \frac{n_{s,g}}{p_s}\sim \alpha_s \frac{n_{s,g}}{p_z} \sim \alpha_s^{3/4} ( Q_s\tau )^{-1/2} Q_s^2,
    \qquad
    T_* \sim\frac{\hat{q}}{\alpha_s m_D^2} \sim\alpha_s^{-3/4} ( Q_s\tau )^{-1/2} Q_s.
\end{align}  
These satisfy the relation $n_{s,g} \sim T_* p_z^2$, consistent with a thermal distribution with temperature $T_*$ up to momenta of order $p_z$.

During this stage, the number of produced quarks remains insufficient to alter the scaling behavior of $\hat{q}$ and $m_D^2$. The quark number density is now given by:
\begin{align}
    n_q \sim 
    \begin{cases} 
          m_D^4 \tau \sim \alpha_s^{3/2} Q_s^3 & \text{for $g\to q/\bar{q}$}\\
          \alpha_s\sqrt{\hat{q}/Q_s} n_{h,g}\tau \sim \alpha_s^{1/2} (Q_s\tau)^{-1/2} Q_s^3 & \text{for $g\to q\bar{q}$}
    \end{cases}
    .
\end{align}  
Here, the dominant contributions for $g\to q\bar{q}$ come from hard gluons. One can also estimate the quark number density from the $g\to q\bar{q}$ splitting of soft gluons:  
\begin{align}
    n_{s,q}\sim \alpha_s\sqrt{\hat{q} /p_s} n_{s,g}\tau\sim \alpha_s\sqrt{\hat{q} /p_z} n_{s,g}\tau \sim \alpha_s^{3/2} Q_s^3.
\end{align}  
This result is parametrically equivalent to that from the $2\leftrightarrow2$ processes and remains negligible compared to the number density of hard quarks until $ Q_s\tau  \sim \alpha_s^{-2}$ when $n_{s,q}\sim n_{q}$. Thus, during this stage, the quark number density follows
\begin{align}
    n_q \sim 
    \begin{cases} \alpha_s^{1/2} ( Q_s\tau )^{-1/2} Q_s^3 & \text{for $\alpha_s^{-3/2}\lesssim  Q_s\tau  \lesssim \alpha_s^{-2}$}\\
    \alpha_s^{3/2} Q_s^3 & \text{for $\alpha_s^{-2}\lesssim  Q_s\tau  \lesssim \alpha_s^{-5/2}$}
    \end{cases}
    .
\end{align}
It is parametrically smaller than the gluon number density until $ Q_s\tau \sim\alpha_s^{-5/2}$ when $n_q\sim n_{h,g}$. Hence, quarks can be neglected in the parametric estimates of $\hat{q}$, $m_D^2$ and other overall properties of the system. Accordingly, the pressure anisotropy scales as $P_L/\epsilon\sim p_z^2/Q_s^2\sim \alpha_s$.

At $ Q_s\tau  \sim \alpha_s^{-5/2}$, the number densities of hard gluons, soft gluons, and quarks become comparable. The soft sector forms a thermalized QGP with temperature $T_* \sim \alpha_s Q_s$, where the relaxation time $t_{\text{rel}} \equiv 1/(\alpha_s^{2} T_*) \sim \tau$, $p_s \sim T_*$, and $n_q \sim n_{s,g} \sim T_*^3$. Similar to the end of the second thermalization stage in spatially homogeneous, very dilute systems, the soft sector reaches its maximum overcooling in terms of $T_*$~\cite{BarreraCabodevila:2022jhi, Cabodevila:2023htm}.

{3. \it Thermalization: heating up a QGP bath via  mini-jet quenching during $\alpha_s^{-{5}/{2}}\lesssim Q_s\tau \lesssim\alpha_s^{-{13}/{5}}$}.
When $  Q_s\tau  \gtrsim \alpha_s^{-\frac{5}{2}} $, the contributions from hard gluons to $ \hat{q} $ and $ m_D^2 $ become parametrically smaller than those from soft gluons and quarks. Additionally, the soft sector can thermalize independently, forming a thermalized QGP with a temperature $ T_* $, since the relaxation time $ t_{\text{rel}} = 1/(\alpha_s^{2} T_*) $ becomes smaller than $ \tau $. As a result, the system undergoes bottom-up thermalization similar to pure gluon systems, except that the soft thermal bath is now a QGP that is made of quarks, antiquarks and gluons. 

More specifically, considering the typical energy loss of each hard parton, $ p_{\text{br}} \sim \alpha_s^2 \hat{q} \tau^2 $ within a time of order $ \tau $~\cite{Baier:2000sb, Baier:2001yt, Blaizot:2013hx, Iancu:2015uja}, the temperature $ T_* $ can be estimated from the energy loss of hard partons: $ \varepsilon_{\text{loss}} \sim T_*^4 \sim p_{\text{br}} n_{h,g} $ with $n_{h,g}\sim Q_s^2/(\alpha_s \tau)$, yielding~\cite{Baier:2000sb}
\begin{align}
T_*\sim  \alpha_s^3 Q_s^2 \tau\Rightarrow \hat{q}\sim \alpha_s^2 T_*^3\sim\alpha_s^{11} Q_s^6 \tau^3,\qquad 
m_D^2\sim \alpha_s T_*^2\sim \alpha_s^7 Q_s^4 \tau^2,\qquad \varepsilon_{\text{loss}}\sim T_*^4\sim\alpha_s^{12} Q_s^{8}\tau^4,
\end{align}
and the number density of soft partons is given by
\begin{align}
    n_{g}\sim n_q \sim T_*^3 \sim \alpha_s^9 Q_s^6 \tau^3.
\end{align}
As the energy is still predominantly carried by hard gluons and the longitudinal pressure is predominantly given by soft partons, the momentum anisotropy scales as $P_L/\epsilon\sim T_*^4/(Q_s n_{h, g})\sim \alpha_s^{13}( Q_s\tau )^5$. Thermalization is completed when hard partons are fully quenched, with
$
\varepsilon_{\text{loss}}\sim Q_s n_{h,g}$,
which gives $ Q_s\tau \sim \alpha_s^{-{13}/{5}}$. At this time, the QGP is heated up to $T_*\sim \alpha^{{2}/{5}} Q_s$ and $P_L/\epsilon\sim 1/3$. All these scalings are the same as those in pure gluon systems~\cite{Baier:2000sb, Wu:2020xtg}.

{4. \it Post-thermalization: Bjorken expansion during $ Q_s\tau \gtrsim\alpha_s^{-{13}/{5}}$.}
After the system thermalizes, it undergoes hydrodynamic expansion with Bjorken flow: $n_g\sim n_q \propto \tau^{-1}$, $T_*\propto \tau^{-1/3}$, $\hat{q}\propto\tau^{-1}$. Starting from their values at $ Q_s\tau \sim\alpha_s^{-13/5}$, we have
\begin{align}
    T_*\sim \alpha_s^{-7/15} Q_s( Q_s\tau )^{-1/3},\qquad n_q\sim n_g \sim \alpha_s^{-7/5} Q_s^2\tau^{-1},\qquad \hat{q}\sim \alpha_s^{3/5} Q_s^2\tau^{-1},\qquad m_D^2 \sim \alpha_s^{1/15} Q_s^2( Q_s\tau )^{-2/3},
\end{align}
as one always has the relaxation time $t_{\text{rel}}=1/(\alpha_s^{2} T_*) \ll \tau$.

\subsection{\bf Quantitative results}

Verifying the scalings  of the various quantities estimated above requires a small coupling, which poses a numerical challenge. Since solving BEDA is computationally more efficient, we concentrate on numerical verifications using BEDA in this section and defer a detailed comparison between BEDA and EKT at intermediate coupling to the next section. 

To numerically solve the BEDA for longitudinally boost-invariant systems, we extend the GPU algorithm from Ref.~\cite{Cabodevila:2023htm}, originally designed for spatially homogeneous systems, by incorporating the dependence on $v_z = p_z / p$ and longitudinal expansion. The numerical simulations are performed on a lattice in $p$ and $v_z$, where the grid spacing must be sufficiently small for accuracy but is constrained by finite computing resources and time. For the results presented in this section, we use a logarithmic grid in $p$ with $N_p = 64$ points, ranging from $p_{\text{min}}/Q_s = 0.01$ to $p_{\text{max}}/Q_s = 12$. The $v_z$ grid is symmetric around $v_z = 0$ and it is distributed around it as a power law $v_{z, i} = x_i^{1.3}$, where $x_i$ is the $i$-th element of a linear grid, and the total number of $v_z$ grid points is set to $N_{v_z} = 64$. With this, the $v_z$ grid is more dense at mid-rapidity, which helps to handle the fact that the distribution functions get very squeezed when the coupling is small. Further details on the implementation of our algorithm will be provided in an upcoming publication~\cite{Cabodevila:2025pzphi}.

To account for the initial momentum anisotropies, we use the same initial distributions as those used in Ref.~\cite{Kurkela:2015qoa}:
\begin{align}
\label{eq:fs_init}
f_0 = \frac{A}{4\pi N_c \alpha_s}\frac{e^{-\frac{2}{3}\left( (p_z \xi)^2 + p_\perp^2 \right) / Q_0^2}}{\sqrt{\left( (p_z \xi)^2 + p_\perp^2 \right) / Q_0^2}},\qquad F=0,
\end{align}
where $Q_0=1.8Q_s$, $A=10.68$ and the initial momentum anisotropy of the system is parametrized by the variable $\xi$. In our results the number of active quark flavors is $N_f = 3$.

\begin{figure}[H]
\centering
\includegraphics[width=0.45\textwidth]{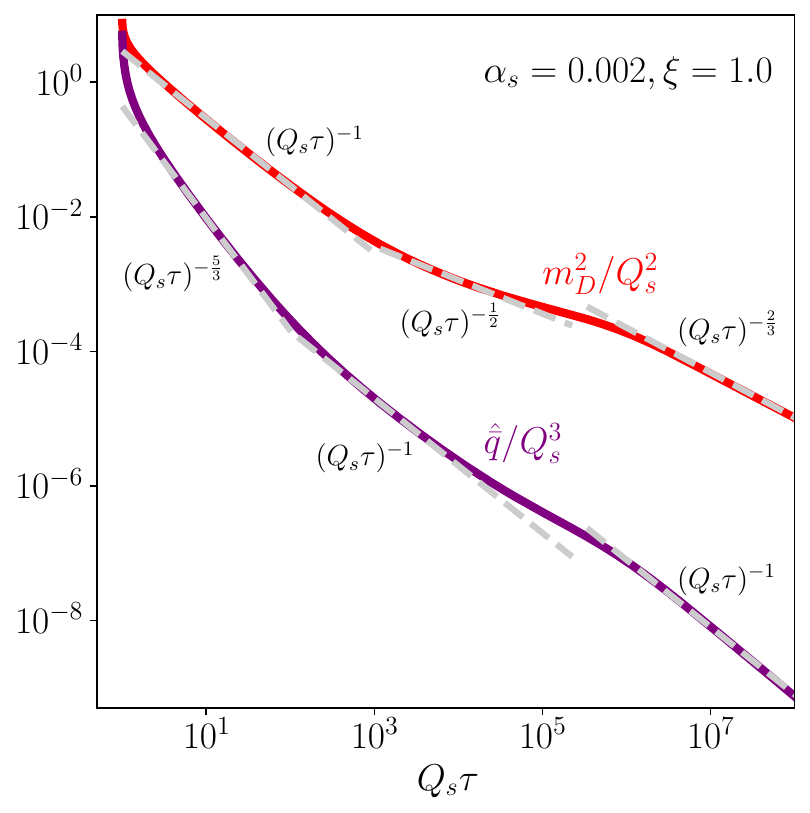}
\hspace{0.05\textwidth}
\includegraphics[width=0.45\textwidth]{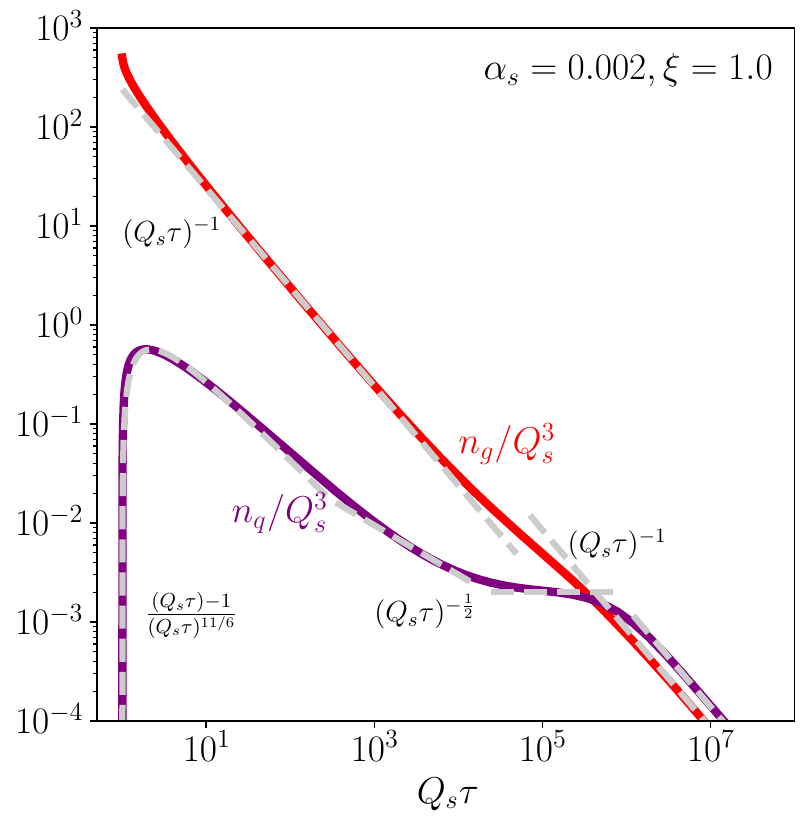}

\includegraphics[width=0.45\textwidth]{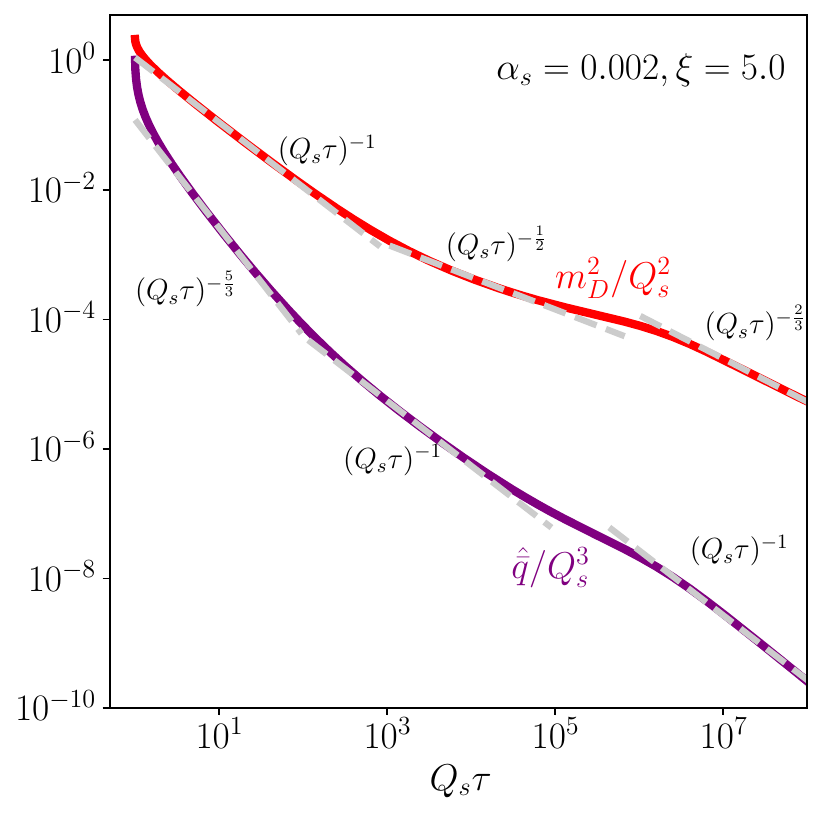}
\hspace{0.05\textwidth}
\includegraphics[width=0.45\textwidth]{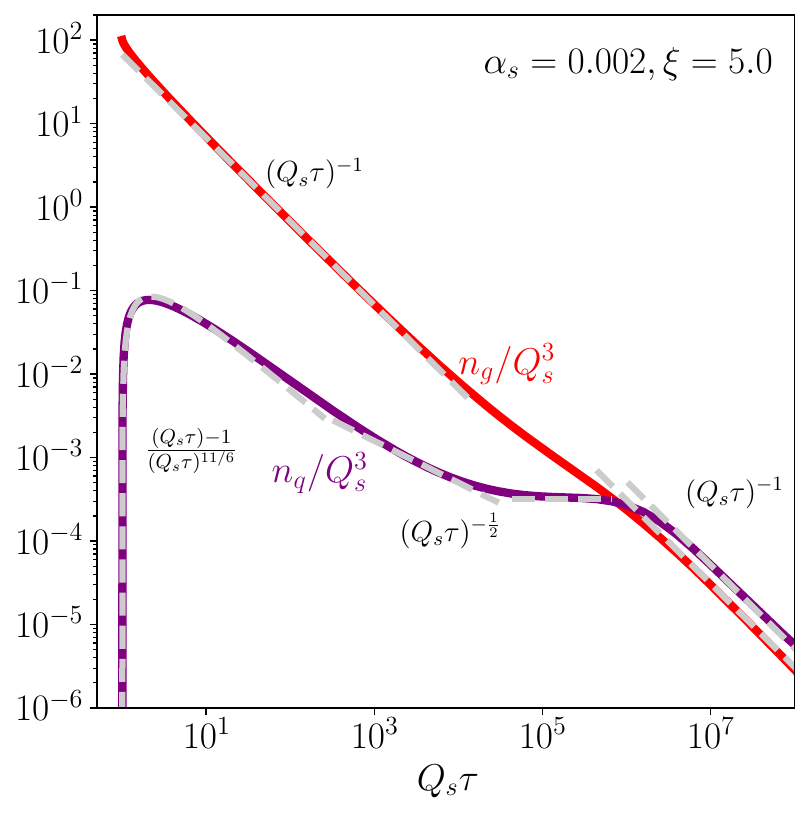}
\caption{
{
Quark production in the bottom-up thermalization at $\alpha_s = 0.002$. The top and bottom panels display $\hat{\bar{q}}$, $m_D^2$, $n_g$ and $n_q$ for the initial anisotropies $\xi =1.0$ and $\xi =5.0$, respectively. Dashed lines represent the estimated scalings based on parametric estimates, excluding those from the third stage.
}
}
\label{fig:scaling_weak_coupling}
\end{figure} 

Figure~\ref{fig:scaling_weak_coupling} shows our simulation results for $\hat{\bar{q}}$, $m_D^2$, $n_g$, and $n_q$ at $\alpha_s = 0.002$, compared with the scalings estimated in the previous subsection. We analyze two sets of initial distributions with $\xi = 1$ and $\xi = 5$. As illustrated by the dashed lines, these quantities approach the predicted scalings during the first two stages and at late times in Bjorken hydrodynamics. Importantly, the quark number density remains lower than the gluon density until the final stage of thermalization, consistent with our parametric estimates, confirming that quark production does not interfere with the bottom-up thermalization process. However, for $\alpha_s = 0.002$, the results do not exhibit the scalings in the third stage when $\hat{\bar{q}}$ and $m_D^2$ increase with $\tau$, which would require simulations at an even smaller coupling.

\begin{figure}
\centering
\includegraphics[width=0.45\textwidth]{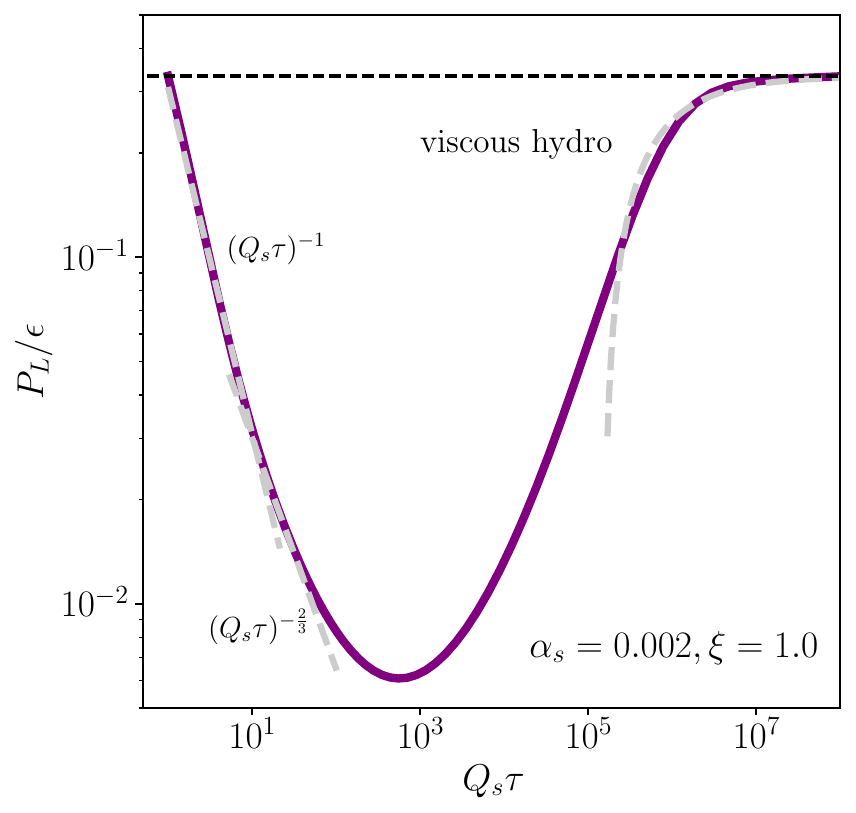}
\hspace{0.05\textwidth}
\includegraphics[width=0.45\textwidth]{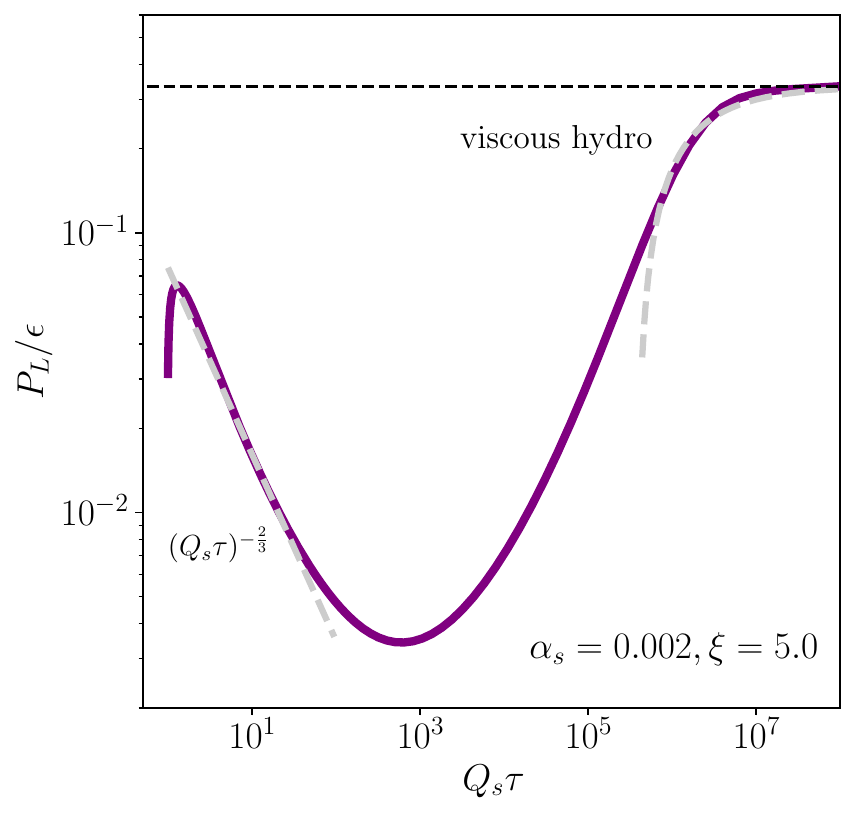}
\caption{
Pressure isotropization and hydrodynamization at $\alpha_s=0.002$. For both initial distributions with $\xi=1.0$ (left) and $\xi=5.0$ (right), our simulations show convergence to the expected scaling at early times as well as hydrodynamization at late times. In both plots, the dashed horizontal line indicates the equilibrium value of 1/3.
}
\label{fig:PLe}
\end{figure}

In Fig.~\ref{fig:PLe}, we further verify the expected scaling in the process of pressure isotropization and hydrodynamization for the same initial conditions as those in Fig.~\ref{fig:scaling_weak_coupling}. For both cases, $P_L/\epsilon$ approaches the expected $\tau^{-2/3}$ scaling~\cite{Baier:2000sb, Schlichting:2012es, Kurkela:2015qoa, Brewer:2019oha, Rajagopal:2024lou} at early times. For the isotropic initial distributions shown in the left panel, the expansion drives the momentum anisotropy initially, leading to a scaling $\sim\tau^{-1}$. In late times, they approach the prediction in viscous hydrodynamics with 
\begin{equation}
    \frac{P_L}{\epsilon} = \frac{1}{3} - \frac{16}{9} \frac{\eta}{s} \frac{1}{T_{hydro} \tau},
\end{equation}
where the ratio of shear viscosity to entropy density, ${\eta}/{s}$, is taken as a fitting parameter to our numerical results at late times when $P_L/\epsilon \gtrsim 0.25$, and $T_{hydro}$ is the temperature given by Landau matching:
\begin{align}
T_{hydro} = \left[ \frac{120 \epsilon}{\pi^2 (4\nu_g + 7N_f \nu_q)} \right]^\frac{1}{4}
\end{align}
with $\nu_g = 16$ and $\nu_q = 6$.

\section{Comparison with EKT at intermediate coupling}

\begin{figure}[ht]
    \centering
    \begin{subfigure}[b]{0.3\textwidth}
        \includegraphics[height=\textwidth]{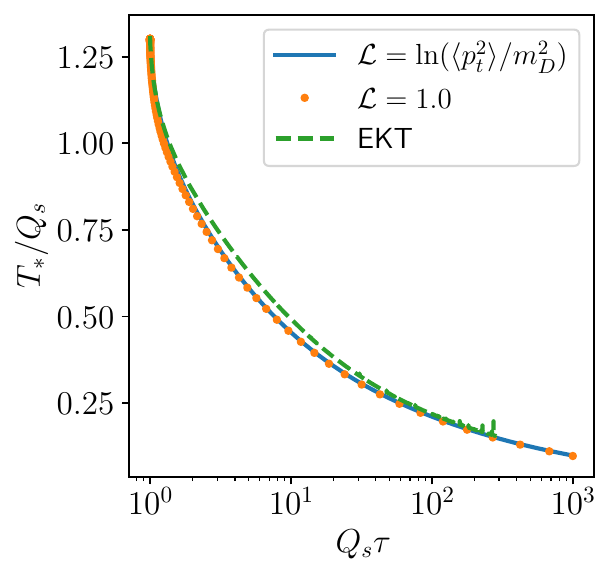}
    \end{subfigure}
    \hfill
    \begin{subfigure}[b]{0.3\textwidth}
        \includegraphics[height=\textwidth]{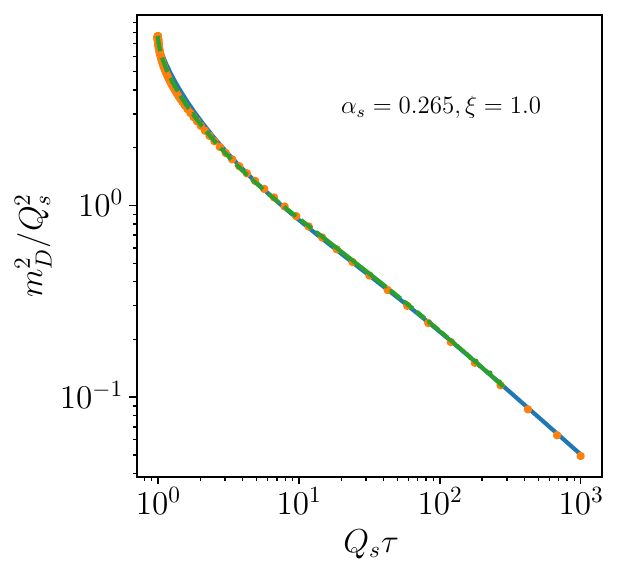}
    \end{subfigure}
    \hfill
    \begin{subfigure}[b]{0.3\textwidth}
        \includegraphics[height=\textwidth]{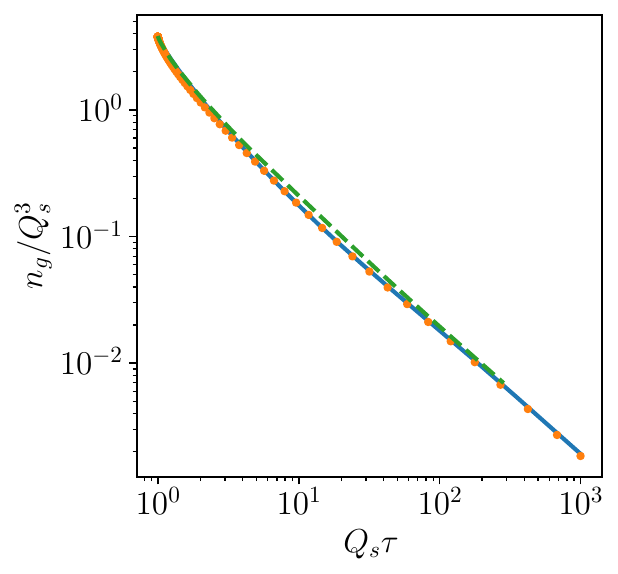}
    \end{subfigure}
    \begin{subfigure}[b]{0.3\textwidth}
        \includegraphics[height=\textwidth]{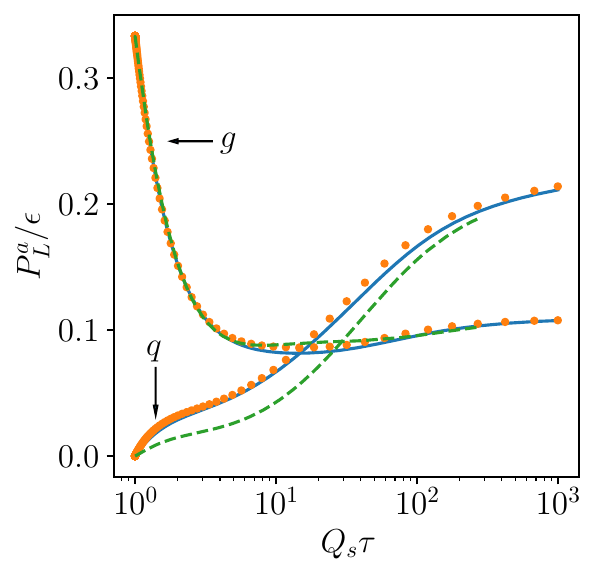}
    \end{subfigure}
    \hfill
    \begin{subfigure}[b]{0.3\textwidth}
        \includegraphics[height=\textwidth]{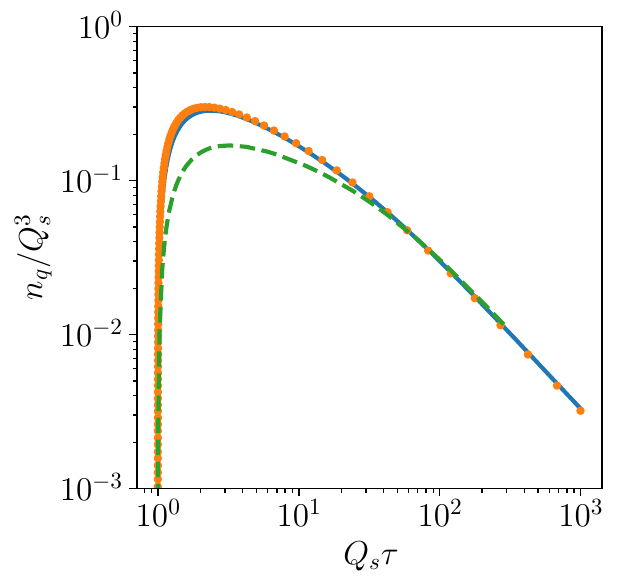}
    \end{subfigure}
    \hfill
    \begin{subfigure}[b]{0.3\textwidth}
        \includegraphics[height=\textwidth]{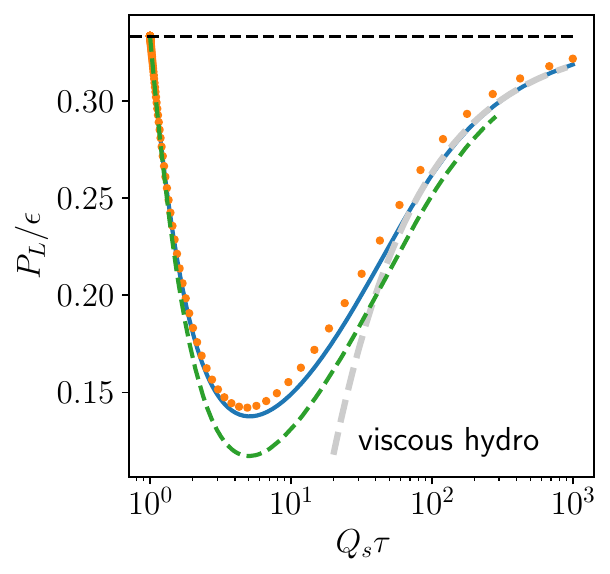}
    \end{subfigure}
    \caption{ Comparison between BEDA and EKT at $\lambda = 10$ ($\alpha_s = 0.265$). The BEDA results with two different choices of $\mathcal{L}$ are presented for various quantities and compared with the EKT results from Ref.~\cite{Du:2020dvp}. The top row displays $T_*$, $m_D^2$, and $n_g$, while the bottom row presents  $P_L^a/\epsilon$, $n_q$ and $P_L/\epsilon$. Here, $P^a_L$ denotes the contribution to the longitudinal pressure from parton species $a$. \label{fig:cmp_with_EKT}
    }
\end{figure}

To quantitatively compare BEDA and EKT, we carry out a detailed analysis at intermediate coupling using the initial distributions given in Eq.~(\ref{eq:fs_init}), as illustrated in Fig.~\ref{fig:cmp_with_EKT}. Here, we set $\xi=1.0$ and $\alpha_s = 0.265$ ($\lambda \equiv g^2 N_c = 10$), a value relevant to heavy-ion phenomenology at LHC energies. For the BEDA results, we use $N_{v_z} = 64$ and $N_p = 64$ with $p_{\text{min}}/Q_s = 0.05$ and $p_{\text{max}}/Q_s = 12$. The EKT results shown in these plots are generated using the code implementation from Ref.~\cite{Du:2020dvp}. For the EKT results, we use a linear grid in $v_z$ with $N_{v_z}=128$ and  the same logarithmic grid in $p$ as BEDA with $N_p=64, p_{\rm min}/Q_s=0.01$ and $p_{\rm max}/Q_s=16$.

In Fig.~\ref{fig:cmp_with_EKT}, we present the BEDA results using two different choices for the logarithmic factor: $\mathcal{L} = 1.0$, as previously chosen in Refs.~\cite{Blaizot:2013lga, Blaizot:2014jna, BarreraCabodevila:2022jhi,Cabodevila:2023htm}, and our new choice in Eq.~(\ref{eq:Lchoice}). All the quantities shown in this figure are not sensitive to these choices of $\mathcal{L}$ at this coupling. Moreover, $T_*$, $m_D^2$, $n_g$ and $P^g_L / \epsilon$ show remarkable agreement with the EKT results over the entire time range, while $P^q_L / \epsilon$, $n_q$ and $P_L / \epsilon$ show some quantitative differences. Here, $P^a_L$ denotes the contribution to the longitudinal pressure from parton species $a$.
For a smaller coupling, such as $\lambda = 5$, we have checked that the quantitative agreement between BEDA with $\mathcal{L}$ from Eq.~(\ref{eq:Lchoice}) remains comparable to that shown in the figure, but the results with $\mathcal{L} = 1.0$ deteriorate. This confirms that our new choice of $\mathcal{L}$ in Eq.~(\ref{eq:Lchoice}) is more consistent than $\mathcal{L}=1.0$ when comparing to the EKT.

The quantitative difference observed in the bottom plots of Fig. \ref{fig:cmp_with_EKT} primarily stems from the difference in quark production. Given that at the initial time there are no quarks in the system, the $1\to2$ process is more efficient for producing quarks and antiquarks than the $2\leftrightarrow2$ process, qualitatively similar to the weak coupling limit in Sec.~\ref{sec:quarkprod}. For the most relevant momentum range, $p\lesssim Q_s$, we find that the deep LPM splitting rates always have a phase space in $x$ that exceeds the corresponding rates in EKT. As a result, this leads to the overestimation of quark production in the BEDA, as shown in the left two plots of the bottom panel in this figure, as well as the observed difference in $P_L/\epsilon$ in the right plot.

Finally, it is intriguing to observe that, even though quark production differs in earlier stages, as the system hydrodynamizes at $Q_s\tau\gtrsim 50$ (see the dashed line in the bottom right panel of Fig. \ref{fig:cmp_with_EKT}), the quark number density in BEDA converges to that in EKT (central bottom panel). This indicates that the system reaches chemical equilibration, as the difference in the gluon number density between BEDA and EKT is remarkably small. That is, the ratio of quark number density to gluon number density converges to the same value in both approaches.

\section{Conclusions}

In summary, we have extended the parametric estimates for bottom-up thermalization in longitudinally boost-invariant systems~\cite{Baier:2000sb} by incorporating quark production in the weak-coupling regime. Our analysis demonstrates that quark production does not modify the three-stage thermalization process observed in pure gluon systems, and the system's overall characteristics remain parametrically consistent with those of pure gluons. Quarks and antiquarks become relevant only in the final stage of thermalization, where they form a QGP thermal bath in equilibrium with soft gluons, quenching hard gluons to complete the thermalization process. Additionally, we have investigated the time-dependent scaling behavior of the quark number density across different thermalization stages.  

Additionally, we have developed a GPU-based algorithm to numerically solve the BEDA for longitudinally boost-invariant systems. Our detailed simulations using BEDA for $\alpha_s = 0.002$ show that the numerical results generally agree with the predicted scaling behaviors from our parametric estimates, except in the third stage of thermalization, which may only become evident at an even smaller $\alpha_s$. A comparison between BEDA and EKT for $\alpha_s=0.265$ demonstrates reasonable agreement, with some noticeable difference in the quark sector, reinforcing BEDA as a valuable and computationally efficient alternative for studying nonequilibrium dynamics in QCD. Future work will focus on extending BEDA simulations to higher dimensions to study thermalization/hydrodynamization in small collision systems.

\section*{Acknowledgements}
This work is supported by the European Research Council project ERC-2018-ADG-835105 YoctoLHC; by Mar\'\i a de Maeztu grant CEX2023-001318-M  and by  project PID2023-152762NB-I00 both funded by MCIN/AEI/10.13039/-501100011033; from the Xunta de Galicia (CIGUS Network of Research Centres) and the European Union.
B.W. acknowledges the support of the Ram\'{o}n y Cajal program with the Grant No. RYC2021-032271-I and the support of Xunta de Galicia under the ED431F 2023/10 project. S.B.C. acknowledges the support of the Axudas de apoio \'a etapa predoutoral program (Ref. ED481A 2022/279).
\bibliographystyle{JHEP}
\bibliography{bulk.bib}

\end{document}